\documentstyle[aps,prl,preprint]{revtex}
\input epsf
\newcommand{\newc}{\newcommand}
\newc{\gsim}{\lower.7ex\hbox{$\;\stackrel{\textstyle>}{\sim}\;$}}
\newc{\lsim}{\lower.7ex\hbox{$\;\stackrel{\textstyle<}{\sim}\;$}}
\def\ms{M_{\rm SUSY}}

\newc{\be}[1]
{\begin{equation} \mbox{$\label{#1}$}}
\newc{\bea}[1]
{\begin{eqnarray} \mbox{$\label{#1}$}}
\newc{\ee}{\end{equation}}
\newc{\eea}{\end{eqnarray}}

\def\re#1{{[\ref{#1}]}}
\def\bm#1{{\mbox{\boldmath $#1$}}}

\begin{document}
\footnotesep=14pt
\begin{titlepage}

\begin{flushright}
\baselineskip=16pt
{\footnotesize
FERMILAB--Conf-97/179-A}\\
{\footnotesize astro-ph/9706271}
\end{flushright}
\renewcommand{\thefootnote}{\fnsymbol{footnote}}
\vspace{0.15in}
\baselineskip=24pt

\begin{center}
{\Large\bf CP-Violating Solitons in the Early
Universe\footnote{\baselineskip=16pt
Talk given by O. T.  at the conference ``Fundamental Physics at the Birth
of the Universe, II''  in Rome, Italy, 19-24 May 1997; contribution to the
proceedings.}}\\
\baselineskip=16pt
\vspace{0.75cm}
\addtocounter{footnote}{1}

{\bf Ola T\"{o}rnkvist\footnote{\baselineskip=16pt
Address after 24 Sept.~1997: DAMTP, Univ.~of
Cambridge, Silver Street, Cambridge CB3~9EW, England.  Email: {\tt
olat@fnal.gov}} and
{\bf Antonio Riotto\footnote{\baselineskip=16pt
PPARC Advanced Fellow,  Oxford Univ. from Sept.~1997.
{}From 1 Dec.~1997 on leave of absence at CERN, Theory Division, CH--1211
Geneva 23,
Switzerland.  Email:
{\tt riotto@fnal.gov}}}
}, \\
\vspace{0.4cm}
{\em NASA/Fermilab Astrophysics Center},\\
{\em Fermi National Accelerator Laboratory},\\
{\em Batavia, Illinois~60510-0500, USA}
\vspace{0.3cm}
\vspace*{0.75cm}

{June 4, 1997}\\
\end{center}
\baselineskip=20pt
\begin{quote}
\begin{center}
{\bf\large Abstract}
\end{center}
\vspace{0.2cm}
{\baselineskip=10pt
Solitons in  extensions of the Standard Model can serve as localized
sources of CP violation. Depending on their stability properties, they may
serve either to create or to deplete the baryon asymmetry.
The conditions for  existence of a particular soliton candidate,  the
membrane solution of the two-Higgs model,  are presented.  In the
generic case, investigated by Bachas and Tomaras, membranes exist and
are metastable for a wide range of parameters.  For the more viable
supersymmetric
case, it is shown that the present-day existence of CP-violating membranes is
experimentally excluded, but preliminary studies suggest that they may have
existed in the early  universe soon after  the electroweak phase transition,
with important consequences for the baryon asymmetry of the universe.}
\vspace*{8pt}


\end{quote}
\end{titlepage}

\newpage
\renewcommand{\thefootnote}{\arabic{footnote}}
\setcounter{footnote}{0}
\baselineskip=16pt
\renewcommand{\baselinestretch}{1.5}
\footnotesep=16pt

In this talk I report on work done in collaboration with Antonio
Riotto\re{CPmem}
on CP-violating  solitons and their possible consequences for the baryon
asymmetry
of the Universe. The talk is organized as follows: After reviewing some
well-known
sources of CP violation in models of particle physics,  I propose that
solitons in
extensions of the Standard Model can serve as {\em localized\/} sources of  CP
violation.
{}From then on I limit the discussion to the type of soliton considered in
Ref.~\re{CPmem} -- membranes.
The conditions for existence of CP-violating
membranes are  presented in two important cases,
corresponding to the smallest possible extensions of the Standard Model:
(1)~a generic model with two Higgs doublets, and (2)~the supersymmetric
two-Higgs-doublet model,
better known as the Minimal Supersymmetric Standard Model (MSSM).
Finally, I discuss the implications of the existence of CP-violating
solitons in the early Universe
and argue that they may, depending on their stability properties,
serve either to create or to deplete the baryon asymmetry.

\subsection*{Sources of CP Violation}

In the Standard Model (SM) of particle physics,  CP non-conserving processes
are the
result of an irreducible complex phase in the
Cabibbo-Kobayashi-Maskawa matrix,
which describes the mixing of  quark flavors of different generations.
Because this phase enters only in higher-loop processes involving
massive charged $W$ bosons, CP violation in the SM is strongly suppressed.
Except in highly speculative scenarios, it cannot account for the observed
baryon asymmetry of  the Universe \re{bau1}.

Extensions of the SM with two Higgs doublets $H_1$ and $H_2$ provide an
additional source of  CP violation in the relative phase $\delta$ of their
expectation
values in the ground state,
\be{minpot}
\langle H_1\rangle = \left(\begin{array}{c} 0\\*[-1ex]
v_1\end{array} \right);
\quad \langle H_2\rangle\equiv
e^{i\delta} \left(\begin{array}{c} 0\\*[-1ex]  v_2 \end{array} \right)\ .
\ee
A non-zero value of $\delta$ arises from  loop corrections to the Higgs
potential that
determines the minimum. At finite temperature, the corrections can be large and
lead
to ``spontaneous'' CP violation \re{noi} as $H_1$, $H_2$ acquire expectation
values in
the electroweak phase transition.  It is generally agreed that extended models
of particle physics provide
sources of  CP-violation that are sufficiently strong for the purpose of
baryogenesis.

\subsection*{Localized Sources of CP Violation}

For the sake of simplicity take $\delta=0$.
Then the ground state (\ref{minpot}) contains no spatially
uniform source of CP violation. However, solitonic excitations exist that
include
a  phase $\delta$ which is non-zero only in a localized region of
space. In such a case, CP violation is confined to the interior of the soliton.

As a first example, consider a texture
[Copeland]\footnote{\label{confref}A single name in brackets refers to that
person's
contribution to these proceedings.}
of the form
\be{texture}
 H_1 = \left(\begin{array}{c} 0\\*[-1ex]
f_1(\bm{x})\end{array} \right);
\quad  H_2 =
U(\bm{x})\left(\begin{array}{c} 0\\*[-1ex]  f_2(\bm{x})\end{array} \right)\ ,
\ee
where $U(\bm{x})\in SU(2)$  and $\pi_3[SU(2)]=\bm{Z}$. A spherical solution of
this
form has been investigated by Bachas et al.~\re{BTT}, who found it to be
unstable
for all parameters of the model.

For the rest of the talk I will concentrate on {\em membranes\/}, which are
wall-like
solutions similar to domain walls. The important difference is that a membrane,
much like a soap film,
separates two regions that are in the identical state,
whereas a domain
wall separates different states. Viewed in a local
rest frame, the flat membrane is a static solution depending only on the
coordinate
$x$ orthogonal to the wall:
\be{memb}
 H_1 = \left(\begin{array}{c} 0\\*[-1ex]
f_1(x)\end{array} \right);
\quad  H_2 =
e^{i\theta(x)}\left(\begin{array}{c} 0\\*[-1ex]  f_2(x)\end{array} \right)\ ,
\ee
where $\theta(-\infty)=0$ and $\theta(\infty)=2\pi$. The phase $\theta$
changes from $0$ to $2\pi$ within a distance synonymous with the thickness of
the
membrane $M_A^{-1}$, where  $M_A$ is the mass of the lightest CP-odd
neutral Higgs boson. Therefore, $e^{i\theta}\neq 1$
inside the membrane while $e^{i\theta}= 1$ outside.

The presence of such a kink solution can be easily
understood by noticing that the Lagrangian of the model contains terms similar
to
those of the sine-Gordon model Lagrangian.  In fact,  using the form
(\ref{memb})
and the approximation
$f_1\approx v_1$, $f_2\approx v_2$ the terms
$ | D_\mu H_1|^2+ |D_\mu H_2|^2  + m_3^{\,2} [H^{\dag}_1 H_2  +
H^{\dag}_2 H_1 ] $ reduce to\footnote{Care must be taken to solve for the gauge
potential in terms of $\theta'$ \re{CPmem}.}
 $v_1 v_2 [ \cos\beta \sin\beta\, {(\theta')}^2+ 2m_3^2
\cos \theta]$. Here
$\tan\beta=v_2/v_1$ is the usual constant parameter defined by the ratio of the
two
expectation values. The corresponding potential  $V\propto -\cos\theta$ is
known to give rise to kink solutions that interpolate from one minimum at
$x=-\infty$ to the adjacent minimum at $x=+\infty$.
The analytic solution  is the sine-Gordon kink,
given by $\theta(x)= 4 \tan^{-1} [\exp{(M_A x)}]$, where
 $M_A^2 = m_3^2/( \cos\beta \sin\beta)$
determines
the characteristic thickness. This solution is identical to the time evolution
$\theta(t)$ of the angle of a  pendulum released at rest from its highest
vertical position
and given a small kick so that it swings through
a full circle within a time $\sim M_A^{-1}$
and comes to rest again asymptotically in its highest position as
$t\to+\infty$.

\subsection*{Conditions for the Existence of Membranes}

Let me first discuss the generic model with two Higgs doublets. In this case
the Higgs sector
contains 9 parameters, and the membrane solutions of this model were
investigated by Bachas and Tomaras \re{bt}. They restricted themselves to a
hyperplane
in
parameter space by fixing 6 of the parameters, including $\tan\beta (=1)$.
For the remaining 3 parameters they picked $M_{h^0}/M_A$,
 $M_{H^0}/M_A$ and  $M_{H^\pm}/M_A$, where $h^0$, $H^0$, $H^+$ and
$H^-$ are the CP-even Higgs bosons of the model, and were able to show that
there exist
{\em classically stable\/} membranes when
$M_{h^0}/M_A\gsim 2$,   $M_{H^\pm}/M_A\gsim 2.2$ and $M_{H^0}/M_A\gsim
g(M_{h^0}/M_A)$, $g$ being a monotonically decreasing function. I  want to
emphasize that membranes thus
exist  in a large region of parameter space,
especially since 6 of the parameters were fixed in this study.

Because these solitons are non-topological, classical stability means that they
constitute a local minimum of the energy, that they are {\em metastable\/}, and
that they
will decay after a finite life-time via the process of  tunneling.

I now proceed to the case of the supersymmetric two-Higgs model, or the MSSM.
This model is, of course, much more attractive from the particle physics point
of
view, since supersymmetry is a major motivation for having two Higgs doublets
in the
first place. Supersymmetry provides ways to solve many of the puzzles of the SM
such as the electroweak symmetry breaking, the problem of quadratic divergences
in
the Higgs mass, and stability of the weak scale under radiative corrections
without
fine-tuning. In addition, supersymmetric particles serve as candidates for dark
matter
[Sadoulet].

Together with Antonio Riotto, I have shown that there exist no membrane
solutions
in the MSSM for any parameters unless one includes one-loop quantum corrections
to the potential \re{CPmem}. These corrections are highly significant
because supersymmetry leads to accidental conspiracies in the classical
tree-level
potential.  In order to address the existence of membranes, one therefore {\em
must\/}
consider the one-loop corrected potential. The low-energy limit of this model
can be described by
three parameters, $M_A$, $\tan\beta$, and $\ms$, while the rest of the
parameters
are constrained by supersymmetry. Here $\ms$ is the scale of supersymmetry
breaking,
typically in the range $M_W\ll \ms\lsim {\cal O}({\rm few})$ TeV.

The issue of existence of membrane solutions hinges crucially on whether
the magnitudes $f_1$ and $f_2$ of eq.~(\ref{memb}) are able to satisfy their
boundary conditions
$f_i=v_i$ at $x=\pm \infty$. This turns out to require a low value of the mass
$M_A$,
corresponding to thick membrane walls. Fig.~1 shows our results \re{CPmem} for
the region of
parameter space
$(M_A, \tan\beta)$ where solutions exist, for one ``low'' and one ``high''
value of $\ms$.
 We cannot choose $\ms$ much higher without reintroducing the need for
fine-tuning.
Also  in the figure are two bounds. The vertical bound
 is from LEP measurements of $e^+ e^-\to h^0 A^0,
h^0 Z$, requiring $M_A\gsim 62.5$ GeV for $\tan\beta>1$.
The horizontal bound
results from assuming that physics is described by a
Grand Unified Theory (GUT) at a scale of about $10^{16}$ GeV.
In such a case
the value of the top Yukawa coupling $h_t$ at low energies is predicted by the
theory,
and combining this with the measured value of the top-quark mass $M_t=(175\pm
6)$ GeV
implies that $\tan\beta\gsim1.1$. This minimal value corresponds to a
gauge-mediated
supersymmetry-breaking mechanism.

{}From Fig.~1 it is apparent that no CP-violating membranes exist in the MSSM
at zero temperature (for which our analysis was done), provided that one
abides with the GUT  hypothesis.

One may now ask oneself whether
 the non-existence of membranes at zero temperature is
in fact an attractive feature, and whether membranes could have existed at
temperatures near that of the electroweak phase transition, $T_c\sim 10^2$ GeV,
and vanished at low temperatures.
Our preliminary investigations show that such
a scenario is possible and would result from large finite-temperature quantum
corrections to $M_A^2$, the parameter that most directly influences the
existence of
the membranes. Taking $M^2_{A, T=0}$ slightly larger than the experimental
limit of
$(62.5)^2$GeV$^2$, the question is whether one can suppress the
high-temperature value
$M^2_{A,T>0}=M^2_{A, T=0}+\delta M^2_{A, T>0}$ by means of a sufficiently
large negative correction $\delta M^2_{A, T>0}$ so as to bring it down to
values
$\lsim 20^2$ GeV$^2$ required for membrane solutions.

The dominant contribution to  $\delta M^2_{A, T>0}$ comes from stop loops and
is given by \re{Brignole} $\delta M^2_{A, T>0} =
\mu A_t h_t^2\,f(\frac{M_Q}{T},\frac{M_U}{T})/(\sin\beta\cos\beta)$, where
\be{ma2corr}
f(x,y)\equiv - \frac{3}{\pi^2}\left[ \frac{J'_+(x^2) - J'_+(y^2)}{x^2-y^2}  +
\frac{\pi}{4}\left( \frac{1}{x+y} - \frac{1}{\tilde{x}+\tilde{y}}\right)
\right]
\ee
and
$J_+(x^2)\equiv \int_0^\infty dt\,t^2 \ln [1 - \exp(-\sqrt{t^2 + x^2})]$.
Here $\mu$, $A_t$, $M_Q$ and $M_U$ are parameters in the stop mass matrix,
and a tilde (\~{}) implies the use of 1-loop self-energy corrected masses
$\tilde{M}_Q$ and $\tilde{M}_U$ \re{Brignole}. For a concrete example,
take  $-\mu=A_t=M_Q=M_U=T=145$ GeV and
$\tan\beta=1$. We then find $f(1,1)=.0433\ldots$, $\delta M^2_{A, T>0} \approx
-3644$ GeV$^2$, and $M^2_{A,T>0} \approx 16^2$ GeV$^2$.  This indicates
that membranes may exist in the MSSM
at $T\lsim T_c$. Riotto and I are currently investigating
this issue more carefully.

\subsection*{Implications for the Baryon Asymmetry}

In electroweak models, the chiral anomaly permits processes at high temperature
that do not conserve the baryon ($B$) and lepton ($L$) numbers, but conserve
$B-L$
\re{bau1}.
These $B$-violating {\em sphaleron\/} transitions involve passing over an
energy
barrier of height $E_{\rm sph}\propto (| H_1|^2 + | H_2|^2)^{1/2}$,
where $E_{\rm sph}\approx 10$ TeV at zero temperature. The $B$-violating
transition probability is proportional to  $\exp(-E_{\rm sph}/T)$.

As we can see from Fig.~2, the magnitudes $f_i=| H_i|$, $i=1,2$, decrease
inside
a membrane, leading to a lower barrier and higher transition rate. Moreover,
our results indicate that $M_A\ll M_W$ in order for membranes to exist. This
means that
the membrane is thick enough to accommodate the field configuration at the top
of
the barrier, the sphaleron, which has a characteristic size of $M_W^{-1}$.

{}From now on let us assume that membrane solutions do exist at high
temperatures.
The picture I want to present features membranes forming inside expanding
critical bubbles in a first-order phase transition [Gleiser]$^1$
at temperatures
just below the critical temperature. The membranes are bounded by string loops,
in analogy with axionic walls [Sakharov], or  (more rarely) form closed
surfaces.
Because the size of a critical bubble is much larger than $M_A^{-1}$,
many membranes  form inside a bubble.

The implications of these membranes are different depending on their stability
properties. If they are metastable, they can move across the plasma and
create non-equilibrium conditions as they strike particles in their way. Since
CP
is violated in the interior of the membrane, and the $B$-violation rate is
enhanced
there,
all three Sakharov conditions [Riotto] may be satisfied, and a net $B (=L)$ can
be generated
that will survive if the phase transition is strongly enough first-order.

If, on the other hand, the membranes are unstable, they will still be created
as
thermal fluctuations inside a critical bubble. The nucleation rate for a
size-$R$ membrane
is of the order of $T \exp[-F(R)/T]$. In order for membranes to play a role,
this rate must be larger than the expansion rate of the universe, $H=T^2/M_{\rm
Pl}$,
which is true if $F(R)/T\lsim 40$. This means that small membranes are
ubiquitous
as fluctuations and provide the conditions for $B$ and CP violation in the
critical
bubble.  Because this scenario occurs in thermal equilibrium, no net $B+L$
can be generated. The membranes can, however,
serve to deplete pre-existing
baryon asymmetry.\footnote{\baselineskip=16pt%
This does not affect the asymmetry in models of GUT
baryogenesis where a net $B-L$ is created at the GUT scale.}

\subsection*{Conclusions}

In summary, I have demonstrated that both the generic and the supersymmetric
two-Higgs Standard Model contain CP-violating soliton solutions. If Nature is
described by a generic two-Higgs model, then there exist metastable membranes
that provide a new mechanism for creating the baryon asymmetry of the universe.
If Nature is supersymmetric, such membranes
are excluded  at zero temperature, but may exist (unstable or metastable)
near the electroweak phase transition temperature.
\begin{figure}[t]
\epsfxsize=15.5cm\epsfbox[18 190 592 460]{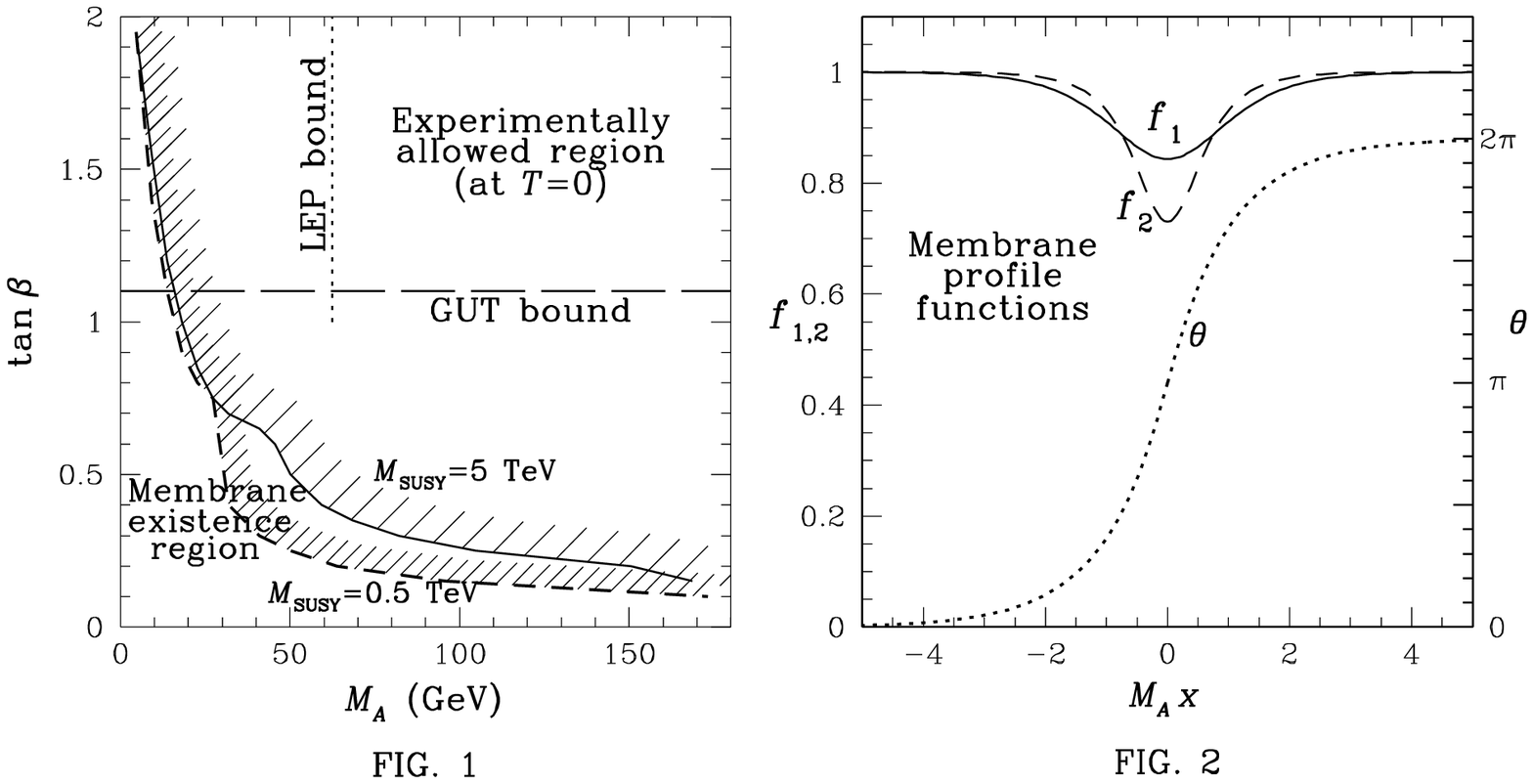}
\end{figure}

I emphasize that CP-violating solitons, whatever their particular realization,
can
play an important role in critical bubbles immediately after a first-order
phase
transition, serving either to create or deplete the baryon asymmetry.

\vspace{-3mm}
\subsection*{ACKNOWLEDGMENTS}

Antonio and I thank F. Occhionero, L. Amendola and C. Baccigalupi for
organizing
a successful conference.
The work of O.T. and A.R. is supported in part by the DOE and NASA
under Grant NAG5--2788. O.T. is also supported
by the Swedish Natural Science Research Council (NFR).

\vspace{-3mm}
\subsection*{REFERENCES}


\def\NPB#1#2#3{Nucl. Phys. {\bf B#1} (19#2) #3}
\def\PLB#1#2#3{Phys. Lett. {\bf B#1} (19#2) #3}
\def\PLBold#1#2#3{Phys. Lett. {\bf#1B} (19#2) #3}
\def\PRD#1#2#3{Phys. Rev. {\bf D#1} (19#2) #3}
\def\PRL#1#2#3{Phys. Rev. Lett. {\bf#1} (19#2) #3}
\def\PRT#1#2#3{Phys. Rep. {\bf#1} (19#2) #3}
\def\ARAA#1#2#3{Ann. Rev. Astron. Astrophys. {\bf#1} (19#2) #3}
\def\ARNP#1#2#3{Ann. Rev. Nucl. Part. Sci. {\bf#1} (19#2) #3}
\def\MPL#1#2#3{Mod. Phys. Lett. {\bf #1} (19#2) #3}
\def\ZPC#1#2#3{Zeit. f\"ur Physik {\bf C#1} (19#2) #3}
\def\APJ#1#2#3{Ap. J. {\bf #1} (19#2) #3}
\def\AP#1#2#3{{Ann. Phys. } {\bf #1} (19#2) #3}
\def\RMP#1#2#3{{Rev. Mod. Phys. } {\bf #1} (19#2) #3}
\def\CMP#1#2#3{{Comm. Math. Phys. } {\bf #1} (19#2) #3}

\def\labelenumi{[\theenumi]}
\begin{enumerate}
\itemsep=0pt
\parskip=0pt
\item\label{CPmem} A. Riotto and O. T\"{o}rnkvist, {\em CP-Violating
Solitons in the Minimal Supersymmetric Standard Model\/}, hep-ph/9704371,
to be published in Phys.\ Rev.\ {\bf D}.

\item\label{bau1}  For a recent review, see for instance, V. A. Rubakov and
M. E. Shaposhnikov, Usp.\ Fiz.\ Nauk. {\bf 166}, 493 (1996).

\item\label{noi} D.\ Comelli, M.\ Pietroni,  Phys. Lett. {\bf B306}
(1993)
67; D.\ Comelli, M.\ Pietroni,  A.\ Riotto, Nucl. Phys. {\bf
B412}
(1994) 441;  J.R.\ Espinosa, J.M.\ Moreno, M.\ Quir\'{o}s,  Phys. Lett. {\bf
B319} (1993) 505.



\item\label{BTT}
C.\ Bachas, P.\ Tinyakov, T.N.\ Tomaras, Phys.~Lett.~{\bf B385} (1996) 237.

\item\label{bt} C.\ Bachas and T.N.\ Tomaras, Phys.\ Rev.\ Lett.\ {\bf 76}
(1996)
356.


\item\label{Brignole}
A.\ Brignole, J.R.\ Espinosa, M.\ Quiros, F.\ Zwirner,
Phys.\ Lett.\ {\bf B324} (1994)181.

\end{enumerate}
\end{document}